\documentclass[showpacs,twocolumn]{revtex4-1} 
\usepackage{braket}
\usepackage[usenames,dvipsnames]{xcolor}

\pdfoutput=1
\usepackage{latexsym,diagbox,stackrel}
\usepackage{mathrsfs}
\usepackage{bbold}
\usepackage{amssymb}
\usepackage{amsmath}
\usepackage{multirow}
\usepackage[T1]{fontenc}
\definecolor{light-gray}{gray}{0.95}

\usepackage{tikz}
\usepackage{graphicx}
\usetikzlibrary{shapes.geometric, arrows}

\usepackage[color=gray]{todonotes} % use with \todo{} or \todo[inline]{} % disable w/ package option [disable]

\newcommand{\C}{\mathbb{C}}

\newcommand{\CP}{\mathbb{CP}}

\newcommand{\R}{\mathbb{R}}

\newcommand{\cI}{\mathcal{I}}

\newcommand{\Z}{\mathbb{Z}}

\newcommand{\dbar}{\bar\partial}
\newcommand{\e}{\mathrm{e}}

\newcommand{\cM}{\mathcal{M}}

\newcommand{\rd}{\, \mathrm{d}}

\newcommand{\be}{\begin{equation}\label}
\newcommand{\ee}{\end{equation}}
\newcommand{\bea}{\begin{eqnarray}\label}
\newcommand{\eea}{\end{eqnarray}}

\begin{document}

\title{Loop Integrands for Scattering Amplitudes from the Riemann Sphere}

\author{Yvonne Geyer$^\dagger$, Lionel Mason$^\dagger$, Ricardo Monteiro$^\dagger$,  Piotr Tourkine$^\ddagger$ \vspace{.1cm}
%\address{
\\ \small{$^\dagger$Mathematical Institute, University of Oxford, Woodstock Road, Oxford OX2 6GG, UK
%The Andrew Wiles Building, The Radcliffe Observatory %Quarter, Oxford OX2 6GG, United Kingdom
\\ % \vspace{.1cm}
$^\ddagger$DAMTP, University of Cambridge, Wilberforce Road, Cambridge CB3 0WA, UK}}
%\pacs{99.9x}
%\keywords{perturbative gauge theory, twistor theory}

\begin{abstract}
The scattering equations on the Riemann sphere give rise to remarkable formulae for tree-level gauge theory and gravity amplitudes. Adamo, Casali and Skinner conjectured  a one-loop formula for supergravity amplitudes based on scattering equations on a torus.   We use a residue theorem to  transform this into a formula on the Riemann sphere.  What emerges is a framework for loop integrands  on the Riemann sphere that promises to have wide application, based on off-shell scattering equations that depend on the loop momentum. We present new formulae, checked explicitly at low points, for supergravity and super-Yang-Mills amplitudes and for  $n$-gon integrands at one loop. Finally, we show that the off-shell scattering equations naturally extend to arbitrary loop order, and we give a proposal for the all-loop integrands for supergravity and planar super-Yang-Mills theory.
\end{abstract}

\maketitle

\section{Introduction}

Worldsheet formulations of  quantum field theories have had wide ranging impact on the study of scattering amplitudes, from the conceptual simplicity of having one basic object instead of multitudes of Feynman diagrams to unexpected new structures such as those relating gauge theory and gravity.  They lie at the heart of string theory, and more recently, in the form of twistor- and ambitwistor-string theory, have been applied to many conventional field theories.  However, the mathematical framework becomes very challenging on the higher-genus worldsheets required to describe loop effects, and it is difficult to see how the relative simplicity of the expected amplitudes arises from integrals involving theta functions over moduli spaces of Riemann surfaces. In this letter, we show how in such worldsheet models  based on the scattering equations, we can transform formulae on higher-genus surfaces to ones on the Riemann sphere. The framework can potentially be applied more generally in field theory.

Given $n$ null momenta $k_i$, the scattering equations determine $n$ points $\sigma_i$ on a Riemann sphere, up to M\"obius transformations.   
The scattering equations not only arise from conventional string theory at low tension \cite{Gross:1987ar}, they also 
underpin the remarkable formulae for tree-level scattering amplitudes of gauge theory and gravity that arise from twistor-string theories \cite{Witten:2004cp,Roiban:2004yf} and the more recent formulae in arbitrary dimension due to  Cachazo, He and Yuan (CHY) \cite{Cachazo:2013hca,Cachazo:2014xea}.  The CHY formulae arise from ambitwistor string theories  \cite{Mason:2013sva,Casali:2015vta,Ohmori:2015sha}, and Adamo, Casali and Skinner (ACS) \cite{Adamo:2013tsa} showed that these  lead to  formulae for 10-dimensional type-II supergravity 1-loop amplitudes in terms of scattering equations on an elliptic curve or torus (and, in principle, to $g$ loops on curves of genus $g$). The ACS 1-loop proposal was investigated further by Casali and one of us \cite{Casali:2014hfa}, motivating the $n$-gon conjecture for an expression on the elliptic curve which gives rise to loop integrands based on permutations of polygons. At fixed loop momentum, both the ACS and the $n$-gon formulae localise to give a sum of residues, involving Jacobi theta functions and the modular parameter $\tau$ for the elliptic curve. Despite the successful factorisation checks, the question remains as to how such formulae could reduce to the rational expressions we expect of loop integrands.

Here we modify the scattering equations on the elliptic curve so as to be able to identify a well-defined loop integrand.  We then use a residue theorem in the modular $\tau$-plane to derive new formulae on a nodal Riemann sphere, without Jacobi theta functions.
%, from the ACS and the $n$-gon conjectures.
 New off-shell scattering equations determine the location of the nodes of the Riemann sphere, where the off-shell loop momentum is inserted. The new formulae give 1-loop integrands which are rational functions of the momenta. This is a scheme that in principle extends to provide formulae for multiloop integrands, applicable also to other theories.

The existence of a canonical loop integrand for non-planar gauge theories and gravity is controversial.  There are many choices for the loop momentum $\ell$, as  we can translate $\ell$ in any particular diagram.  The formulae we obtain give a natural global choice, although not an obvious one from the perspective of Feynman diagrams. In order to transform standard integrands into our formulae, we must perform a variety of shifts.  

In this paper, we first prove the equivalence of the ACS and $n$-gon conjecture on elliptic curves to corresponding formulae on the Riemann sphere, giving new 1-loop proposals.  These pass various checks and the general form of the loop integrand that arises is given.  We also propose a formula on the Riemann sphere for the 1-loop integrand of super Yang-Mills, and subject it to similar checks.
 Finally, we give a brief discussion of the  extension to an all-loop conjecture for type-II supergravity and super Yang-Mills theory.  At $g$-loops, these loop integrands have the same level of complexity as $n+2g$-point trees.  Further details will appear elsewhere.

%The off-shell formulations we obtain are no longer subject to the stringent anomaly cancellation conditions of conventional string theory nor ambitwistor-string theories.  It will be interesting to explore analogous phenomena in conventional string theory.

\section{Scattering equations on a torus}
We use the complex coordinate $z$ on the elliptic curve $\Sigma_q =\C/\{\Z\oplus \Z \tau\}$ where $q=\e^{2\pi i \tau}$.
The scattering equations are equations for $n$ points $z_i\in \Sigma_q$ that depend on $n$ momenta $k_i \in \R^d$, $i=1,\ldots n$.  To define them we construct a meromorphic 1-form $P(z,z_i|q)d z$ on $\Sigma_q$ that satisfies
$$
\dbar P=2\pi i \sum_i k_i \bar \delta(z-z_i) dz\, , 
$$
where
$
\bar \delta (f(z)):= \dbar \frac{1}{2\pi i f(z)} =\delta(\Re f)\delta(\Im f) d\overline{f(z)}\, .
$
Introducing $\ell\in\R^d$ to parametrise the zero-modes, and denoting $\;\;z_{ij}=z_i-z_j$, our choice of solution for $P(z,z_i|q)$ is
\be{P-def}
P%(z;z_i)
=2\pi i\,\ell d z +  \sum_i k_i \Bigg( \frac{\theta'_1 (z-z_i)}{\theta_1 (z-z_i)} + \sum_{j\neq i}  \frac{\theta'_1 (z_{ij})}{n\, \theta_1 (z_{ij})} \Bigg)dz\, .
\ee
This is meromorphic and doubly periodic
%\footnote{that in  \cite{Adamo:2013} is not meromorphic and that in \cite{Casali:2014hfa} is not doubly periodic in the $z_i$.} 
in $z$ \emph{and} the $z_i$.  The ACS version is not holomorphic and does not factorise properly \cite{Adamo:2013tsa}, while that in \cite{Casali:2014hfa,Ohmori:2015sha} is not doubly periodic until  the loop momentum is integrated out (as in conventional string theory), and is thus not well defined on the elliptic curve for fixed $\ell$.
Using \eqref{P-def}, the scattering equations are
\be{SE}
\mathrm{Res}_{z_i} P^2(z)=2k_i\cdot P(z_i)=0\, ,  \qquad P^2(z_0)=0\, .
\ee
%where we note that $P^2(z_0)$ is independent of  $z_0$ 
  Because the sum of residues of $P^2$ vanishes, the first scattering equation follows from those at $i=2,\ldots, n$. Translation invariance implies that we must fix the location of $z_1$ by hand. On the support of the equations at $z_i$, which fix these points, $P^2(z_0)$ is global and holomorphic, hence constant in $z_0$, depending only on $\tau$. Therefore, the final equation $P^2(z_0)=0$ determines $\tau$.

The ACS proposal for the 1-loop integrand of type-II supergravity amplitudes takes the form
\be{elliptic-amp}
\cM^{(1)}%{\mathrm{1-loop}}
_{\mathrm{SG}}%(1,\ldots, n)
=\int \cI_q %(k_i,\epsilon_i,z_i,q)
\,d^d \ell \,  d\tau \, \bar\delta (P^2(z_0))\prod_{i=2}^n \bar\delta(k_i\cdot P(z_i)) d z_i \,,
\ee
where, for the critical case, $d=10$ and $\cI_q=\cI(k_i,\epsilon_i, z_i|q)$, and $\epsilon_i$ is the polarisation data. It is obtained as a sum over spin structures of a worldsheet correlator of vertex operators, giving rise to certain Pfaffians and partition functions described later and in more detail in \cite{Adamo:2013tsa}.   This formula is doubly periodic in the $z_i$ and modular invariant, i.e., invariant  under $\tau\rightarrow \tau+1, -1/\tau$ (and $\ell\rightarrow \ell,\tau\ell$).

In \cite{Casali:2014hfa}, it was shown that when $n=4$, as in string theory, $\cI$ is independent of $z_i$ and $q$, so it factors out of the integral.  The nontrivial remaining integral is the $n=4$ version of the more general integral
$$
\cM^{(1)}_{n\mathrm{-gon}}=\int \,d^{d} \ell \,  d\tau \, \bar\delta (P^2(z_0))\prod_{i=2}^n \bar\delta(k_i\cdot P(z_i)) d z_i \,,
$$
where the integral is modular invariant for $d=2n+2$.  In \cite{Casali:2014hfa},  this was conjectured to be equivalent to a sum over permutations of $n$-gons.

In both cases, there are as many delta functions as integration variables and these restrict the integral to a sum over a discrete set of solutions to the scattering equations.  Each term consists of the integrand evaluated at the corresponding solution divided by a Jacobian.  

\section{From a torus to a Riemann sphere}
Here we use a residue  theorem (or integration by parts in our notation) to reduce the formula on the elliptic curve to one on the nodal Riemann sphere at $q=0$ (such `global residue theorems' have already been applied to tree-level CHY formulae by  \cite{Dolan:2013isa}).   
We will be left with scattering equations that have  off-shell momenta associated to $\ell$, and a formula for the 1-loop integrand based on these.

\vspace{-.2cm}
\begin{figure}[h]
    \includegraphics[width=0.44\textwidth]{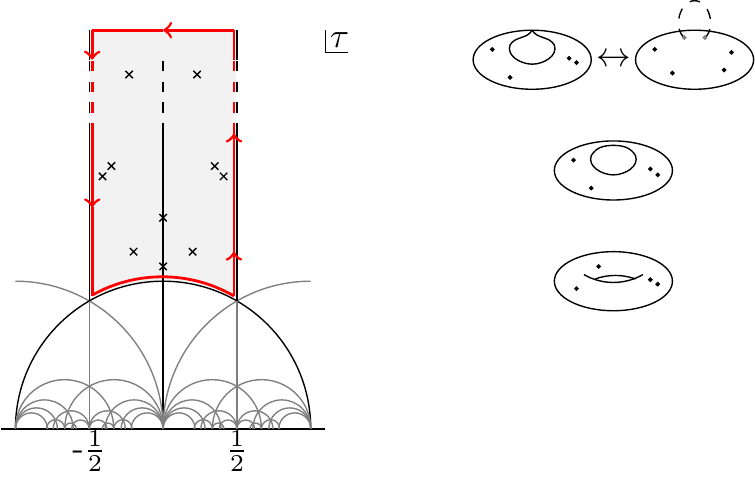}
\caption{Contour argument in the fundamental domain.}
\label{fund-dom}
\end{figure}

In order to obtain a formula for the amplitude on the Riemann sphere, we assume that $\cI_q:=\cI(\ldots|q)$ is holomorphic as a function of $q$ on the fundamental domain $D_\tau=\{|\tau|\geq 1, \Re \tau \in[-1/2,1/2]\}$ for the modular group. It was shown in \citep{Adamo:2013tsa} that the holomorphicity of the supergravity integrand at $q=0$ is a consequence of the GSO projection. For other values of $q$ the possible poles in the integrand can only occur when $z_i\rightarrow z_j$, but the standard factorisation argument \cite{Dolan:2013isa} applies here also to imply that this can only happen when the momenta are  factorising and hence nongeneric. The main argument is then
\begin{eqnarray}
\cM^{(1)}_{SG}&=&\int \cI_q\,d^d \ell \,  \frac{d q}{ q} \, \dbar \left(\frac{1}{2\pi i P^2(z_0)}\right)\prod_{i=2}^n \bar\delta(k_i\cdot P(z_i)) d z_i \nonumber \\
&=& -\int \cI_q\,d^d \ell \,  \dbar\left(\frac{d q}{2\pi i q} \right)\,  \frac{1}{ P^2(z_0)}\prod_{i=2}^n \bar\delta(k_i\cdot P(z_i)) d z_i \nonumber \\
&=& -\int \cI_0\,d^d \ell \,    \frac{1}{  P^2(z_0)}\prod_{i=2}^n \bar\delta(k_i\cdot P(z_i)) d z_i \,
\Big|_{q=0}\, .
\end{eqnarray}
In the first line, we put $d\tau=dq/2\pi i q$ and inserted the definition of $\bar\delta (P^2(z_0))$.  In the second line, we integrated by parts in the domain $D_\tau$, yielding a delta function supported at $q=0$ that is then integrated out.  The boundary terms cancel because of the modular invariance.  This is equivalent to a contour integral argument in the fundamental domain $D_\tau$ as in figure \ref{fund-dom}. The sum of the residues at the poles of $1/P^2(z_0,\ldots |q)$ simply gives the contribution from the residue at the top, $q=0$, since the contributions from the sides and the unit circle cancel by modular invariance.  

The fundamental domain for $z$ maps,
\begin{equation}
\sigma= \e^{2\pi i (z-\tau/2)}\,,
\end{equation}
 to $\{\e^{-\pi\Im \tau} \leq|\sigma|\leq\e^{\pi\Im \tau}\}$, with the identification $\sigma\sim q \sigma$. As  $q\rightarrow0$, we obtain $\sigma\in\CP^1$ with $0,\infty$  identified, giving a double point corresponding to the pinching of $\Sigma_q$ at a non-separating degeneration. We have $dz=\frac{d\sigma}{2\pi i\sigma}$ and
$$
 \frac{\theta'_1 (z-z_i)}{\theta_1 (z-z_i)}dz=\frac{\pi}{\tan \pi (z-z_i)} \,dz =-\frac{d\sigma}{2\,\sigma} + \frac{d\sigma}{\sigma-\sigma_i}\, ,
$$
 at $q=0$, so using momentum conservation we obtain 
\begin{equation}
P(z)=P(\sigma)= \ell \,\frac{d\sigma}{\sigma} +\sum_{i=1}^n \frac{k_i\, d\sigma}{\sigma-\sigma_i}\, ,
\end{equation}
where here we have translated $\ell$ by $\sum_{i\neq j}k_i \cot \pi z_{ij}$. %(by the time we are on the Riemann sphere there are no periodicity issues to worry about although the shift does change the degree of the scattering equations).

If we now consider the function $P^2(\sigma)$, we find that it has double poles at $0$, $\infty$ (along with the usual simple poles at $\sigma_i$).  Defining $S=P^2-\ell^2 \rd \sigma^2/\sigma^2$, we find $S$ now has only simple poles. The vanishing of the residues of $S$ gives our {\em off-shell} scattering equations 
\be{SE2n}
0=\mathrm{Res}_{\sigma_i}S =k_i\cdot P(\sigma_i) = \frac{k_i\cdot \ell}{\sigma_i} + \sum_{j\neq i}\frac{k_i\cdot k_j}{\sigma_i-\sigma_j}\, , 
\ee
at $\sigma_i$.  The sum of the residues of $\sigma_\alpha\sigma_\beta S$ must vanish with $\sigma_\alpha=(1,\sigma)$ in affine coordinates, so that the equations for $i=2,\ldots , n$ imply the vanishing of the residues of $S$ at $\sigma_1$, $0$ and $\infty$. Thus any $n-1$ of these equations imply all $n+2$, hence $S$ is holomorphic and, having negative weight, vanishes, so that $P^2=\ell^2\rd \sigma^2/\sigma^2$.

With this, the 1-loop formula becomes
\be{1-loop}
\cM^{(1)}_{SG}= -\int \cI_0\,d^d \ell \, \frac{1}{\ell^2}\prod_{i=2}^n \bar\delta(k_i\cdot P(\sigma_i))\frac{d \sigma_i}{\sigma_i^2}\, ,
\ee
where we have used the identity $\bar\delta(\lambda f)=\lambda^{-1}\bar\delta(f)$ to give $\bar{\delta}(k_i\cdot P(z_i)) dz_i=
\bar{\delta}(k_i\cdot P(\sigma_i))d\sigma_i/\sigma_i^2$. The formula \eqref{1-loop} is our new proposal for the supergravity loop integrand, with $\cI_0$ the $q=0$ limit of the ACS correlator.

For the simpler `$n$-gon' conjecture presented in \cite{Casali:2014hfa}, we now take $\cI_q=1$. For both this and supergravity, modular invariance is no longer an issue on the Riemann sphere, and the new formulae make sense in any dimension.

\section{Shifts and the $n$-gon conjecture}
When $n=4$, the $n$-gon conjecture implies the supergravity conjecture.  Here the off-shell scattering equations can be solved exactly with two solutions,  %This problem is identical to that arising in factorisation as studied in \cite{Casali:2014hfa} except that now $\ell$ is off-shell.  It was there conjectured there to have $(n-1)!-2(n-2)!$ solutions and at $n=4$ the two solutions can be found explicitly 
yielding
\begin{equation*}\label{4pt-result}
\hat{\cM}^{(1)}_4 = \frac{-1}{\ell^2} \sum_{\sigma \in S_4} 
\frac{1}{\ell\cdot k_{\sigma_1}(\ell\cdot (k_{\sigma_1}+ k_{\sigma_2})+k_{\sigma_1}\cdot k_{\sigma_2})\ell\cdot k_{\sigma_4}}\,.
\end{equation*}
Henceforth we denote $\cM=\int \hat{\cM} \,d^d \ell$.
This is not obviously equivalent to the permutation sum of the boxes
\begin{equation}
  \label{eq:Ibox}
I^{1234} = \frac{1}{\ell^2(\ell+k_3)^2(\ell+k_3+k_4)^2(\ell-k_2)^2}\,,
\end{equation}
%.  The corresponding integrand is
%\begin{equation}
%B_4 = \frac12 (I^{1234} + I^{1243} + I^{1324} + I^{4321} + I^{3421} + I^{4231}),
%\end{equation}
%where The factor $1/2$ in $B_4$ is due to the fact that we are symmetrising with respect to $\ell\to-\ell$, and therefore sum over six boxes, rather than three.
as the only manifest propagator in $\cM^{(1)}_4$ is the pre-factor ${1}/{\ell^2}$, and all the other denominator factors are linear in $\ell$. 
% \cdot K$, where $K$ is a combination of the external momenta. To match the analytic structure of $B_4$, we must perform shifts of the loop momentum.  
However,  the partial fraction identity
\begin{equation}
  \label{eq:Di-partfrac}
  \frac1{\prod_{i=1}^{n}D_{i}} =
  \sum_{i=1}^{n}\frac{1}{D_{i}\prod_{j\neq i}(D_{j}-D_{i})}
\end{equation}
can be applied to a contribution such as \eqref{eq:Ibox}.   The right-hand-side of this identity is a sum of terms with a single factor of the type $D_i=(\ell+K)^2$, and several factors of the type $D_{j}-D_{i}=2\ell\cdot K +\mathcal{O}(\ell^0)$. We then perform a shift in the loop momentum for each term such that the corresponding $D_i$ is simply $\ell^2$. Applying this procedure to the permutation sum, we precisely obtain $\hat{\cM}_4^{(1)}$.

More generally, using the above expansion and shifts, the $n$-gon integral is equivalent to
%\begin{equation}
%\label{eq:conjngon}
$$
\hat{\cM}^{(1)}_{n} %= (-2)^n \,\text{Shift}[n\text{-gons}] 
=\frac{(-1)^n}{\ell^2} \sum_{\sigma\in S_n} \,
\prod_{i=1}^{n-1} \frac{1}{ \ell\cdot\sum_{j=1}^i k_{\sigma_i} +\frac{1}{2} \big(\sum_{j=1}^i k_{\sigma_i}\big)^2 }\, .
$$
%\end{equation}
That this arises from \eqref{1-loop} with $\cI=1$  has been checked directly numerically at five and six points.  Furthermore,  the locations of the poles of this formula are singled out by an adaptation of a standard worldsheet factorisation argument \cite{Dolan:2013isa}  for the scattering equations \eqref{SE2n}. We remark  
that at two and three points we obtain zero identically, as expected from dimensional regularisation.

\section{Supergravity at 1-loop}
For supergravity, $\mathcal{I}(k_i,\epsilon_i,z_i|q)$  was given in detail in \cite{Adamo:2013tsa} as $\mathcal{I} = \mathcal{I}^L\,\mathcal{I}^R$, with each $\cI^{L/R}$ given as a sum over spin structures of worldsheet correlation functions on $\Sigma_q$,  as in superstring theory. For brevity, we consider only the even spin structures, which suffice if all kinematic data is in dimension 9 or less.
 While $\cI^{L/R}$ are finite as $q\rightarrow 0$, some constituents vanish and others diverge. We obtain the $q=0$ result by extracting the following coefficients
%The final result can be expressed as
\begin{equation}\label{Pfaff0}
%\mathcal{I}^L_0=16\left(\text{Pf}(M_2)-\text{Pf}(M_3)\right)
% -2\,\partial_{q^{1/2}}\text{Pf}(M_3) \,\Big|_{q=0} \, ,
\mathcal{I}^L_0=16\left(\text{Pf}(M_2)-\text{Pf}(M_3)\right) \big|_{q^0}
 -2\,\text{Pf}(M_3) \,\big|_{q^{1/2}} \, ,
\end{equation}
where 
%\begin{equation}
%\mathcal{I}^L=Z_2\,\text{Pf}(M_2) -Z_3\,\text{Pf}(M_3) +Z_4\,\text{Pf}(M_4) ,
%\end{equation}
%where $Z_\alpha=\theta_\alpha(0|\tau)^4/\eta(\tau)^{12}$ are partition functions, $\alpha$ labels the spin structure and  
$\text{Pf}(M_\alpha)$ are  the Pfaffians of the $2n\times 2n$ matrices
\begin{equation}
M_\alpha = \left( \begin{matrix}
A \;\;& -C^T \\
C  \;\;&  B
\end{matrix} \right),
\end{equation}
with $A_{ij} = k_i\cdot k_j \,S_\alpha(z_{ij})$,  $B_{ij} = \epsilon_i\cdot \epsilon_j \,S_\alpha(z_{ij})$,   $C_{ij} |_{i\neq j} = \epsilon_i\cdot k_j \,S_\alpha(z_{ij})$, and
%\begin{equation}
$C_{ii} = -2\pi i\,\epsilon_i\cdot P(z_i)$.
%\end{equation}
The Szeg\"o kernels $S_\alpha$ are defined in \cite{Adamo:2013tsa}.
%\begin{equation}
%S_\alpha(z_{ij}|\tau) = \frac{\theta_1'(0|\tau)}{\theta_1(z_{ij}|\tau)}\,\frac{\theta_\alpha(z_{ij}|\tau)}{\theta_\alpha(0|\tau)}\, \sqrt{dz_i}\sqrt{dz_j}.
%\end{equation}
The substitution $\epsilon_i^\mu\to\tilde\epsilon_i^\mu$ gives $\mathcal{I}^R$, where the supergravity states have polarisation tensors $\epsilon_i^\mu\tilde\epsilon_i^\nu$.
In terms of $\sigma$ we have to ${\mathcal O}(q)$
\begin{align}
S_2(z_{ij}) &\to \frac{1}{2}\,\frac{\sigma_ i+\sigma_ j}{\sigma_ i-\sigma_ j} \sqrt{\frac{d\sigma_ i}{\sigma_ i}} \sqrt{\frac{d\sigma_ j}{\sigma_ j}},
 \nonumber \\ 
S_3(z_{ij}) &\to \left(\frac{\sqrt{\sigma_ i}\sqrt{\sigma_ j}}{\sigma_ i-\sigma_ j}+q^{1/2} \frac{\sigma_ i-\sigma_ j}{\sqrt{\sigma_ i}\sqrt{\sigma_ j}}\right) \sqrt{\frac{d\sigma_ i}{\sigma_ i}}\sqrt{\frac{d\sigma_ j}{\sigma_ j}}. \nonumber
\end{align}

For $n=4$, $\mathcal{I}$ is a constant for any $q$, giving the expected $t_8t_8R^4$ kinematic tensor \cite{Casali:2014hfa}, and the $n$-gon results above suffice to give the correct answer. For $n=5$,  the integrand $\mathcal{I}_0$ depends on the $\sigma_ i$ and the loop momentum. The amplitude can be written in terms of pentagon and box integrals, and we can apply the shift procedure above to connect to our results, yielding 
\begin{align*}
\label{eq:5ptint}
& \hat{\cM}_5^{(1)}=\frac{1}{32\,\ell^2} \sum_{\sigma\in S_5}
\frac{1}{\prod_{i=1}^4 \big(\ell\cdot\sum_{j=1}^i k_{\sigma_i} +\frac{1}{2} (\sum_{j=1}^i k_{\sigma_i})^2\big)} 
\nonumber \\
& \times 
\left( N^5_{\sigma,\ell} \,+ 
\frac{1}{2} \sum_{i=1}^4 N^{\text{box}}_{\sigma_i\sigma_{i+1}} \,\frac{\ell\cdot\sum_{j=1}^i k_{\sigma_i} +\frac{1}{2} (\sum_{j=1}^i k_{\sigma_i})^2}{k_{\sigma_i}\cdot k_{\sigma_{i+1}}} \,   \right).
\end{align*}
The supergravity numerators $N^5$ and $N^{\text{box}}$ are the square of the gauge theory numerators given in \cite{Carrasco:2011mn} or \cite{Mafra:2014gja}, which satisfy the colour-kinematics duality \cite{Bern:2008qj,Bern:2010ue}.  This formula precisely matches that from the off-shell scattering equations at 5 points numerically.

\section{Super Yang-Mills at 1-loop
}
Following CHY at tree level, we can hope to obtain super Yang-Mills amplitudes at 1-loop by replacing one of the factors $\cI^R$ for supergravity above by a cyclic sum over Parke-Taylor factors that run through the loop
\begin{equation}
PT_n=\sum_{i=1}^n \frac{\sigma_{0\, \infty}}{\sigma_{0\, i}\sigma_{i\, i+1}\sigma_{i+1\, i+2}\ldots \sigma_{ i+n\,\infty}}\, ,
\end{equation}
where $\sigma_0=0$ and $\sigma_\infty=\infty$, and $i$ is defined mod $n$. Thus planar 1-loop amplitudes are given by
$$
\hat{\cM}^{(1)}(1,\ldots,n)=\frac{1}{\ell^2}\int {\cI^L_{0}\, PT_n}  \prod_{i=2}^n \bar{\delta}(k_i\cdot P(\sigma_i)) \frac{d\sigma_i}{\sigma_i} \,,
$$
where $\cI^L_0$ is given in \eqref{Pfaff0}. Each factor of $1/\sigma_i$ goes with a Pfaffian, so in removing a Pfaffian, we also remove one of the $1/\sigma_i$s.
 At four points, $\cI^L_0$ is constant as mentioned above and it factors out.
This ansatz has been checked numerically at both four and five points.
There is an additional obvious conjecture for the analogue of the biadjoint scalar theories.
%, which are non-supersymmetric.
%$$
%\cM(1,\ldots,n)=\int PT_n \widetilde{PT}_n \prod_{i=2}^n \bar\delta(k_i\cdot P(\sigma_i))d\sigma_i \, ,
%$$
%where the two different Parke-Taylor factors can be based on different orderings.  This has also been checked numerically????

\section{All-loop integrands}
The ACS proposals have natural extensions to Riemann surfaces $\Sigma^g$ of arbitrary genus $g$ for $g$-loop amplitudes \cite{Adamo:2013tsa,Ohmori:2015sha, Adamo:2015hoa}.  We can again attempt to use residue theorems to localise on a preferred boundary component of the moduli space. Here we choose a basis of $g$ $a$-cycles to contract in $g$ non-separating degenerations, to obtain Riemann spheres  $\Sigma^g_0$ with $g$ nodes, i.e  pairs of double points $(\sigma_r,\sigma_{r'})$, $r=1,\ldots,g$. (We still expect separating degenerations to be suppressed by the remaining scattering equations for generic momenta.) This fixes $g$ of the moduli, and the remaining $2g-3$ moduli are now associated with the $2g$ new marked points modulo M\"obius transformations.  On nodal curves, 1-forms are allowed to have simple poles at the nodes so that the nodal Riemann sphere $\Sigma^g_0$ is endowed with a basis of $g$ global holomorphic 1-forms 
\begin{equation}
 \omega_r=\frac{(\sigma_r-\sigma_{r'})d\sigma}{(\sigma-\sigma_r)(\sigma-\sigma_{r'})}\, .
\end{equation}

For $P$ with poles at $n$ further marked points $\sigma_i$ and residues $k_i$, we have 
\begin{equation}
P=\sum_{r=1}^g \ell_r\omega_r+\sum_i k_i \frac{d\sigma}{\sigma-\sigma_i} \, ,
\end{equation}
where $\ell_r\in \R^d$ are the zero modes in $P$ representing the loop momenta. Setting
$
S(\sigma):=P^2-\sum_{r=1}^g \ell_r^2 \omega_r^2\, ,
$
 a quadratic differential with simple poles at all the marked points including $\sigma_r,\sigma_{r'}$,  the multiloop off-shell scattering 
 equations are
\begin{equation}
\mathrm{Res}_{\sigma_i} S=0\, , \quad i=1,\ldots ,n+2g \,,
\end{equation}
where $i$ now ranges over all the marked points.   We have as before three relations between the scattering equations arising from the vanishing of the sum of the residues of $\sigma_\alpha \sigma_\beta S$.  Thus  if we impose $n+2g-3$ of them, the remaining ones must also be satisfied, so that $S$ is holomorphic and, being of negative weight, vanishes.

This leads to the following proposal for the all-loop supergravity integrand
\begin{equation}
\hat{\cM}^{(g)}_{SG}=\int_{(\CP^1)^{n+2g}} \frac{\cI^L_0\cI^R_0}{\mathrm{Vol}\, G} \prod_{r=1}^{g} \frac{1}{\ell_r^2} \prod_{i=1}^{n+2g} \bar{\delta}(\mathrm{Res}_{\sigma_i} S(\sigma_i))\, ,
\end{equation}
where $G=SL_2\times \C^3$ is the residual gauge symmetry of the ambitwistor string. It is fixed in standard Faddeev Popov fashion by fixing three points to $(0,1,\infty)$ and removing their corresponding delta functions. In this formula, the integrand factors $\cI^R$ and $\cI^L$ depend on the marked points, momenta and polarisation data, and take values in 1-forms in each integration variable. They are most simply defined to be  the sum over spin structures of the worldsheet correlator \cite{Adamo:2013tsa,Adamo:2015hoa} of $n$ type-II supergravity vertex operators on a genus $g$ Riemann surface $\Sigma^g$, then taken to the $g$-fold nodal limit $\Sigma_0^g$.  There is of course much work to be done to make such correlators explicit, but they are the same as those that arise in conventional string theory and as such are much studied. This is done in our context at four points and two loops in \cite{Adamo:2015hoa}.

Similar conjectures can be made for planar super Yang-Mills and for the analogue of bi-adjoint scalar amplitudes at all loops.  We should respectively replace one or both of the $\cI^R$ and $\cI^L$ by  the sum of all Parke-Taylor factors that are compatible with some given ordering of the external particles, but which also  run through all the loops, generalizing the 1-loop case. Indeed, one can conceive of conjectures based on the ingredients of \cite{Cachazo:2014xea,Casali:2015vta} for the theories described there. 

\bigskip

\noindent
{\bf Acknowledgments:}
We thank Oliver Schlotterer and David Skinner for useful discussions. YG is supported by the EPSRC and the Mathematical Prizes fund, LM is partially supported by EPSRC grant number EP/J019518/1, RM is a Marie Curie Fellow and a JRF at Linacre College, and PT is supported by the ERC Grant 247252 STRING.

\bibliography{twistor-bib}  

\end{document}